\documentclass[twoside,twocolumn,english,superscriptaddres,showpacs,longbibliography,aps,prl,preprintnumbers]{revtex4-2}

\usepackage[T1]{fontenc}
\usepackage[utf8]{inputenc}
\usepackage[english]{babel}
\usepackage{bm}
\usepackage{amsmath}
\usepackage{amssymb}
\usepackage{graphicx}
\usepackage[unicode=true, pdfusetitle, bookmarks=true,bookmarksnumbered=true,bookmarksopen=true,bookmarksopenlevel=1,
breaklinks=false,pdfborder={0 0 0},pdfborderstyle={},backref=false,color links, citecolor=blue,linkcolor=red,urlcolor=blue] {hyperref}
\usepackage[nameinlink]{cleveref}
\Crefname{figure}{Fig.}{}

\makeatletter
\def\set@firstnote#1{%
 \@ifnum{\firstnote@num=#1\relax}{}{%
  \class@warn@end{Endnote numbers changed: rerun LaTeX}%
 }%
 \immediate\write\@mainaux{%
   \global\mathchardef\string\firstnote@num#1\relax
 }%
}%

\usepackage{braket}
\usepackage{bm}
\usepackage{tikz}
\usepackage{pgfplots}
\usepackage{pgf}
\usepackage{subfigure}
\usepackage{diagbox}
\usepackage{orcidlink}

\renewcommand{\selectlanguage}[1]{}

\makeatother

\pgfplotsset{compat=1.18}
\begin{document}

\title{Pairing and modular: Generalized electrons and quasiholes inside the $\boxtimes$ of topological orders}
\author{Yoshiki Fukusumi}
\affiliation{Physics Division, National Center for Theoretical Sciences, National Taiwan University, Taipei 106319, Taiwan}
\author{Yuma Furuta}
\affiliation{Institute for Advanced Study, Kyushu University, Fukuoka 819-0395, Japan}
\preprint{KYUSHU-HET-391}
\date{\today}
\begin{abstract}
 We study the relationship between the pair creation mechanism in physics and the pairing ($\boxtimes_{\mathbf{A}_{\text{ub}}}$) in mathematics, focusing on the modular and algebraic properties. We provide a way to couple two topologically ordered systems, $\mathbf{A}$ and $\mathbf{A'}$ sharing the common (non-group-like) structure $\mathbf{A}_{\text{ub}}=\overline{\mathbf{A'}_{\text{ub}}}$, through the nonabelian anyon analog of the pair creation mechanism denoted as $\mathbf{A}_{\text{ub},\boxtimes}=\mathbf{A}_{\text{ub}}\boxtimes \mathbf{A'}_{\text{ub}}$. We identify $\mathbf{A}_{\text{ub},\boxtimes}$ as the generalized version of electron operators in general topological orders or hidden current with (half-)integer spin involving the conformal embeddings. We provide a way to construct $\mathbf{A}_{\text{ub},\boxtimes}$ as the hidden topological operator and identify $\mathbf{A}\boxtimes_{\mathbf{A}_{\text{ub}}}\mathbf{A'}$, a general class of topoloical orders, as the $\mathbf{A}_{\text{ub},\boxtimes}$-extended symmetry topological field theory. We also conjecture a series of underlying (unexplored) modular invariants, connecting the conformal embedding and the pairing.

\end{abstract}

\maketitle

\section{introduction}
\label{introduction}

\emph{Pair} creation is among the most fundamental ideas in physics. The literature and the underlying contexts are vast. Hence, we only list the Bardeen–Cooper–Schrieffer (BCS) pair creation mechanism\cite{Bardeen:1957mv} and the quark-confinement problem\cite{Wilson:1974sk}. In mathematics, from different backgrounds, the corresponding structure has been studied commonly as the \emph{balanced} structure under the action of the transparent subalgebra $\mathbf{A}_{\text{ub}}=\overline{\mathbf{A'}_{\text{ub}}}$ which appears commonly in two (fusion) algebras $\{\alpha\}=\mathbf{A}$ and $\{ \alpha'\}=\mathbf{A'}$. The $\mathbf{A}_{\text{ub}}$ balanced tensor product of $\mathbf{A}-\mathbf{A'}$ bimodule\cite{Kitaev:2011dxc}, $\otimes_{\mathbf{A}_{\text{ub}}}$, and the $\mathbf{A}_{\text{ub}}$ balanced relative Deligne product\cite{Etingof:2009yvg,Douglas2014TheBT}, $\boxtimes_{\mathbf{A}_{\text{ub}}}$, are the structures that we focus. This kind of balanced structure has also been called \emph{pairing} in mathematics. Hence, the ideas of pairing appear commonly in physics and mathematics, and we attempt to strengthen the bridge.

\begin{figure}[htbp]
\begin{center}
\includegraphics[width=0.4\textwidth]{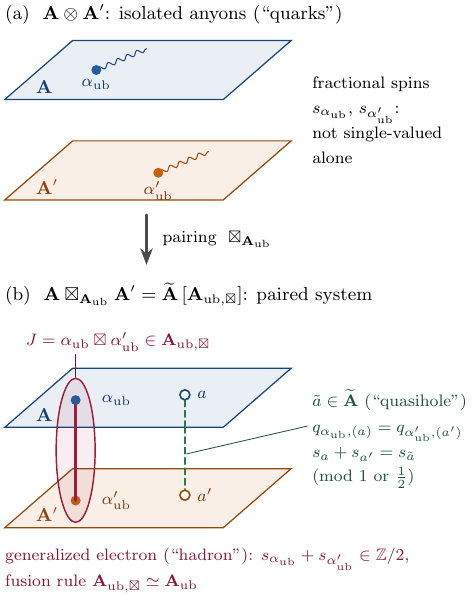}
\caption{Picture of the pairing phenomenon. (a) The indivusual systems have fractional spin particles or quasiholes $\mathbf{A}_{\text{ub}}=\overline{\mathbf{A'}_{\text{ub}}}$ before the pairing. (b) After the pairing, the new electron objects $\mathbf{A}_{\text{ub},\boxtimes}$ appear and the quasiholes are modified to $\widetilde{\mathbf{A}}$ following the anomaly matching condition by $\{ q\}$. The total system becomes $\mathbf{A}\boxtimes_{\mathbf{A}_{\text{ub}}} \mathbf{A'}=\widetilde{\mathbf{A}}[\mathbf{A}_{\text{ub},\boxtimes}]$.}
\label{fig:electron_quasihole}
\end{center}
\end{figure}

We summarize the phenomenology from the conventional or \emph{inside-the-box} view in this paper (FIG.\ref{fig:electron_quasihole}). The underlying mechanism of the pairing $\otimes_{\mathbf{A}_{\text{ub}}}$ is simple only by gluing two modules $\mathcal{H}_{\mathbf{A}}$ and $\mathcal{H}_{\mathbf{A'}}$ (or Hilbert spaces of matters) canonically under the operations $\mathbf{A}_{\text{ub}}$ (isolated charged anyons\cite{Leinaas:1977fm,Goldin:1979ki,Wilczek:1981du,Goldin:2022gcn} or ``quarks"). This structure $\otimes_{\mathbf{A}_{\text{ub}}}$ connects two systems in an organized way through the relation $\mathbf{A}_{\text{ub}}=\overline{\mathbf{A}_{\text{ub}}}$ and this reproduces the anomaly matching\cite{Lieb:1961fr,Schultz:1964fv,tHooft:1979rat,Cho:2017fgz} (or charge conservation). The operation taking $\boxtimes_{\mathbf{A}_{\text{ub}}}$ is more involved. This operation introduces composite algebras $\mathbf{A}_{\text{ub},\boxtimes}=\mathbf{A}_{\text{ub}}\boxtimes \mathbf{A'}_{\text{ub}}$(BCS-like paired anyons or ``hadrons") which can be regarded as the generalization of \emph{electron operators}, generating the single valued wavefunction\cite{Laughlin:1983fy,Moore:1991ks}. Following this straightforward observation, the other objects in $\mathbf{A}\boxtimes_{\mathbf{A}_{\text{ub}}}\mathbf{A'}$, $\widetilde{\mathbf{A}}$, should be regarded as \emph{quasiholes}. For readers in condensed matter, we stress that 
\begin{equation}
\begin{split}
&\text{$\mathbf{A}_{\text{ub},\boxtimes}$ is nonabelian anyonic particles with} \\
&\text{\emph{(half-)integer spin} boson-fermion statistics.}
\end{split}
\end{equation}
The resultant fusion rule is unconventional, but we demonstrate its ubiquity as elementary (but hidden) building blocks of topological orders (TOs)\cite{Wen:1989iv,Wen:1989zg}.

The Verlinde formula\cite{Verlinde:1988sn} is fundamental in constructing a meaningful algebra, fusion rule, in physics. The fusion rule of anyons is determined by their module ($\mathcal{H}$) and modular data ($S$). There exist many established data in the studies of modular forms\cite{Cappelli:1987xt,Christe:1988vc,Gannon:1994km}. When taking $\mathbf{A}=\mathbf{A}_{\text{ub}}=\overline{\mathbf{A'}}$, the structure $\mathbf{A}\boxtimes \overline{\mathbf{A}}$, called spherical fusion category (SFC) or nonchiral fusion ring\cite{Fuchs:1993et,Rida:1999xu,Rida:1999ru,Fukusumi:2024cnl,Nivesvivat:2025odb}, corresponds to modular invariants \cite{Fuchs:2002cm,Kong:2017hcw} with a close connection to the Moore-Seiberg-data\cite{Moore:1988qv,Moore:1988ss}. Hence,  we attempt to continue this research direction to $\mathbf{A}\boxtimes_{\mathbf{A}_{\text{ub}}}\mathbf{A'}$, by providing a systematic way to relate this pairing problem to the conformal embeddings\cite{Schellekens:1986mb} from the coset construction\cite{Goddard:1984vk,Goddard:1986ee,Goddard:1988md} and the level-rank duality\cite{Kuniba:1990zh,Nakanishi:1990hj,Hsin:2016blu,Cordova:2017vab,Cordova:2023jip}. 

The pairing structure in conformal field theories (CFTs) and topological quantum field theories (TQFTs) has been noticed in various research fields. For example, the vertex operator $\Phi_{\alpha_{\text{ub}}\boxtimes \alpha'_{\text{ub}}}(\in\mathbf{A}_{\text{ub},\boxtimes})$ has been studied as \emph{(half-)integer} spin hidden current involving the conformal embeddings\cite{Crnkovic:1989ug,Lashkevich:1993fb,Gaiotto:2014kfa,Furuta:2025ahl,Zhang:2026gqp,Fukusumi:2026pkr} with a connection to massless renoramlization group (RG) flows\cite{Zamolodchikov:1987ti,Zamolodchikov:1989hfa}. When restricting our attention to group-like pairing or simple current extension\cite{Schellekens:1989uf,Schellekens:1990ys,Gato-Rivera:1990lxi,Kreuzer:1993tf}, the operator identification is the related structure, even in textbooks\cite{DiFrancesco:1997nk,Recknagel:2013uja}. Hence, $\mathbf{A}_{\text{ub},\boxtimes}$ can be called (half-)integer spin nonsimple current\cite{Fuchs:2000gv}. In TQFT, related orbifolding has been studied in \cite{Cordova:2023jip,Cordova:2025eim}. More mathematically, the pairing structure has been formulated by the connected \'{e}tale algebra\cite{davydov2011structure,DavydovMugerNikshychOstrik+2013+135+177} or condensable algebra\cite{Kong:2013aya} and plays the central role in the anyon condensation\cite{Bais:2008ni} or proliferation\cite{Cheng:2026qax}. We note a few recent works \cite{Huston:2022utd,Zhang:2024bye,Gannon:2026ttf} in a similar research direction. However, a unified understanding of the pairing $\boxtimes_{\mathbf{A}_{\text{ub}}}$ has not been established. This partially comes from the different research directions in CFTs or TQFTs: 
\begin{itemize}
\item{In CFT, the main scope tends to be the analytical (or local) aspect of $\mathbf{A}_{\text{ub},\boxtimes}$. The global structure of Hilbert spaces ($\otimes_{\mathbf{A}_{\text{ub}}}$) has rarely been studied.}
\item{In TQFT, the main scope tends to be the algebraic (or global) aspect of $\otimes_{\mathbf{A}_{\text{ub}}}$.  The fusion rule of $\mathbf{A}\boxtimes_{\mathbf{A}_{\text{ub}}}\mathbf{A'}$ and the underlying wavefunctions ($\mathbf{A}_{\text{ub},\boxtimes}$) have rarely been studied.}
\end{itemize}
We also emphasize that the identification of the hidden current $\mathbf{A}_{\text{ub},\boxtimes}$ as the generalization of the electron is less familiar. Related ideas can only be seen in \cite{Kong:2019cuu,Kaidi:2021gbs,Fukusumi:2022xxe,Fukusumi_2022_c,Zhang:2024bye,Fukusumi:2024ejk,Fukusumi:2025ljx,Fukusumi:2026pkr} and older studies in fractional quantum Halls\cite{Milovanovic:1996nj,Cappelli:1996np,Ino:1998by,Cappelli:2010jv,Schoutens:2015uia}. Most of the other quantum field theoretic studies focus on the structure of the \emph{orbifolded} theory, $\widetilde{\mathbf{A}}=\left(\mathbf{A}\boxtimes _{\mathbf{A}_{\text{ub}}}\mathbf{A'}\right)/_{\text{ob}} \mathbf{A}_{\text{ub},\boxtimes}$. However, the \emph{extended} theory $\mathbf{A}\boxtimes _{\mathbf{A}_{\text{ub}}}\mathbf{A'}(=\widetilde{\mathbf{A}}\times_{\text{ad}} \mathbf{A}_{\text{ub},\boxtimes})$ is more fundamental in TOs and TQFTs from their wavefunctions.

In this paper, we provide a general construction of $\mathbf{A}\boxtimes _{\mathbf{A}_{\text{ub}}}\mathbf{A'}$, consistent with the conformal embeddings, by applying the modular properties. The main proposals are Eq.\eqref{main_orbifold} and Eq.\eqref{main_extension}. We apply our method to the pairing between the tricritical Ising models and the Ising model in Sec. \ref{section_example} and summarize concluding remarks and open problems in Sec. \ref{section_conclusion}. In the end matter, we introduces further examples and discuss implications of our methods for the existing classifications in CFTs and TQFTs.

\section{$\mathbf{A}\boxtimes_{\mathbf{A}_{\text{ub}}} \mathbf{A'}$ from modular data}
\label{section_main}
\subsection{Construction of paired module $\mathcal{H}\otimes_{\mathbf{A}_{\text{ub}}}\mathcal{H}_{\mathbf{A'}}$}

Let us assume there exist two modules of chiral conformal field theories, i.e., Hilbert spaces of topologically ordered systems, $\{|a\rangle \}=\mathcal{H}_{\mathbf{A}}$ and $\{|a'\rangle \}=\mathcal{H}_{\mathbf{A'}}$ where $\{\alpha\}=\mathbf{A}$ and $\{\alpha'\}=\mathbf{A'}$ are the fusion rings (or sets of anyons) sharing the common subalgebra $\{\alpha_{\text{ub}}\}=\{\overline{\alpha'_{\text{ub}}}\}=\mathbf{A}_{\text{ub}}$. For simplicity, we focus on the chiral-chiral pairing, but one can obtain chiral-antichiral pairing only by replacing $\mathbf{A'}$ with $\overline{\mathbf{A'}}$. The chiral-antichiral pairing has a more direct connection to the Higgs phenomena\cite{Bando:1987br,Moore:1989yh,Cordova:2025eim,Cordova:2023jip}. We also introduce topological operators $\mathcal{Q}$ acting on the modules, such as $\mathcal{Q}_{\alpha}|a\rangle$. We take $\alpha$ and $a$ as labels of primary states in conformal field theories, but we primarily use the Greek symbol $\alpha$ as a label of topological operators and the alphabet $a$ as a label of modules. In the subsequent discussions, we sometimes drop the summation indices and $\otimes$ to simplify the notation. We denote $\mathbf{A}$ as both $\{\alpha \}$ and $\{ \mathcal{Q}_{\alpha}\}$, to focus on the algebraic properties and symmetry actions accordingly. We leave elementary details of CFTs to several textbooks and reviews\cite{Ginsparg:1988ui,DiFrancesco:1997nk,Schweigert:2000ix,Northe:2024tnm}.

We focus on the diagonal model with the modular invariant $\sum_{\alpha}|\chi_{\alpha}(t)|^{2}$, where $\chi$ is the character and $t$ is the modular parameter. In this setting, the action of topological operators (or categorical symmetries\cite{Ji:2019jhk,Kong:2020cie,Chatterjee:2022kxb})\cite{Komargodski:2020mxz,Lin:2022dhv,Fukusumi:2025ljx,Fukusumi:2026tii} is,
\begin{equation}
\mathcal{Q}_{\alpha}|a\rangle=q_{\alpha,(a)}|a\rangle,\  q_{\alpha,(a)}=\frac{S_{\alpha,a}}{S_{I,a}}
\end{equation} 
where $I$ is the vacuum and $S$ is the modular $S$ matrix defined by the modular $S$ transformation $\chi_{a}(-1/t)=\sum_{\beta}S_{a,b}\chi_{b}(t)$, and $q$ is the algebraic generalized quantum dimensions (AGQDs)\cite{Fukusumi:2025xrj,Fukusumi:2025fir}. When $\mathbf{A}_{\text{ub}}$ is a group, $q$ reproduces the chiral 't Hooft anomaly\cite{tHooft:1979rat,Fuchs:1997af,Schweigert:2000ix,Closset:2024sle} or Lieb-Schultz-Mattis-type anomaly\cite{Lieb:1961fr,Schultz:1964fv,Furuya:2015coa,Lecheminant:2015iga,Cho:2017fgz,Yao:2018kel,Tanizaki:2018xto}. By changing the chirality of operators and modules, one can introduce $\mathcal{Q}^{\text{op}}$ and $\overline{\mathcal{Q}}$. The operators can be expressed as,
\begin{equation}
\mathcal{Q}_{\alpha}=\sum_{a} q_{\alpha,(a)}e_{(a)}, \mathcal{Q}^{\text{op}}_{\alpha}=\sum_{a} \overline{q_{\alpha,(a)}}e_{(a)}, \overline{\mathcal{Q}_{\overline{\alpha}}}=\sum_{a} \overline{q_{\alpha,(a)}}\overline{e_{(a)}}
\end{equation}
where $e_{(a)}$ is idempotent satisfying the relation $e_{(a)}\times e_{(b)}=\delta_{a,b}, \ e_{(a)} |b\rangle=\delta_{a,b}|b\rangle$. $\delta$ is the standard Kronecker delta. One can relate the idempotent to the chiral projection $e_{(a)}=\sum_{L} |a, L\rangle \langle a,L|$, where $L$ is the label of the descendant field. The expression $\mathcal{Q}$ and $q$ reproduce the fusion rule,
\begin{equation}
\alpha \times \beta=\sum_{\gamma}N^{\gamma}_{\alpha,\beta} \gamma
\end{equation}
by replacing $\alpha$ to $\mathcal{Q}_{\alpha}$ or $q_{\alpha, (\ ) }$, where $N$ is the fusion coefficients. This is ensured by the Verlinde formula\cite{Verlinde:1988sn}, $N^{\gamma}_{\alpha,\beta}=\sum_{\eta}\left(S_{\eta\alpha}S_{\eta\beta}\overline{S_{\eta\gamma}}\right)/S_{I,\eta}$.

$\mathbf{A}\boxtimes_{\mathbf{A}_{\text{ub}}}\mathbf{A'}$, should be defined by the \emph{balanced} action on the paired module $\mathcal{H}_{\mathbf{A}}\otimes_{\mathbf{A}_{\text{ub}}}\mathcal{H}_{\mathbf{A'}}$. In our setting with $\mathbf{A}_{\text{ub}}=\overline{\mathbf{A'}_{\text{ub}}}$, the paired module is,
\begin{equation}
\begin{split}
&\mathcal{H}_{\mathbf{A}}\otimes_{\mathbf{A}_{\text{ub}}}\mathcal{H}_{\mathbf{A'}} \\
&=\{ |x_{\otimes}\rangle \in \mathcal{H}_{\mathbf{A}}\otimes \mathcal{H}_{\mathbf{A'}} : \ \mathcal{Q}_{\alpha_{\text{ub}}}|x_{\otimes} \rangle=\mathcal{Q}^{\text{op}}_{\alpha'_{\text{ub}}}|x_{\otimes}\rangle)\}
\end{split}
\end{equation}
where $\mathcal{Q}^{\text{op}}$ is introduced from the folding trick\cite{Wong:1994np}.
This pairing chooses a particular combination of the tensor product of modules and constructs a subspace of $\mathcal{H}_{\mathbf{A}}\otimes \mathcal{H}_{\mathbf{A'}}$. When choosing $|x_{\otimes}\rangle=|a\rangle\otimes |a'\rangle$, the pairing condition is $\{q_{\alpha_{\text{ub}},(a)}=\overline{q_{\alpha'_{\text{ub}},(a')}}\}_{\alpha_{\text{ub}}\in\mathbf{A}_{\text{ub}}}$ and relates $\mathcal{Q}^{\text{op}}$ and $\overline{\mathcal{Q}}$. The importance of this pairing has been noticed in \cite{Gaiotto:2012np} based on the module structure studied in \cite{Crnkovic:1989ug}.

The paired module naturally selects the idempotents, and we identify the resultant linear subspace as the weaker version of the Deligne product $\mathbf{A}\boxtimes_{\mathbf{A}_{\text{ub}}}\mathbf{A}$,
\begin{equation}
(\mathbf{A}\boxtimes_{\mathbf{A}_{\text{ub}}}\mathbf{A})_{\text{w}}=\{e_{(a)}e_{(a')}\in \mathbf{A}\otimes \mathbf{A'}:|a\rangle\otimes|a'\rangle\in \mathcal{H}_{\mathbf{A}}\otimes_{\mathbf{A}_{\text{ub}}}\mathcal{H}_{\mathbf{A'}} \}
\end{equation}
where the idempotents above are generators of a semisimple algebra. One can straightforwardly implement the corresponding functor $\boxtimes_{\mathbf{A}_{\text{ub}}} :\mathbf{A}\otimes \mathbf{A}\rightarrow (\mathbf{A}\boxtimes_{\mathbf{A}_{\text{ub}}}\mathbf{A'})_{\text{w}}$ only by applying the projection to each primary state and each idempotent. For quantum states, the operation is a standard diagonal projection from $\mathcal{H}_{\mathbf{A}}\otimes \mathcal{H}_{\mathbf{A'}} $ to $\mathcal{H}_{\mathbf{A}}\otimes_{\mathbf{A}_{\text{ub}}}\mathcal{H}_{\mathbf{A'}} $, choosing Hilbert subspace satisfying the pairing condition $\{q_{\alpha_{\text{ub}},(a)}=\overline{q_{\alpha'_{\text{ub}},(a')}}\}_{\alpha_{\text{ub}}\in\mathbf{A}_{\text{ub}}}$. By focusing on the label of idempotents $e_{(a)}e_{(a')}$, one can construct the corresponding projection to the idempotents. Conventionally, one can construct the $\alpha\boxtimes_{\mathbf{A}_{\text{ub}}}\alpha'$ only by writing down the tensor product $\alpha \otimes \alpha'$ as a linear sum of idempotents and choosing terms satisfying the pairing condition. However, it is uncertain whether the structure satisfies some canonical properties in other formalisms. In the next subsection, we study this point.

We note that we concentrate on the case \emph{without a $\mathbf{A}_{\text{ub}}$ fixed point}. For example, $\mathbb{Z}_{2}$ of the Ising CFT, $\{ I,\psi, \sigma\}$, is not in the main scope. The Ising anyon $\sigma$ is invariant under the application of the $\mathbb{Z}_{2}$ simple current $\psi$, and the system cannot be graded by $\mathbb{Z}_{2}$. This is inconvenient when focusing on the $\mathbb{Z}_{2}$ pairing problem described by a modular tensor category (MTC).  

\subsection{Construction of orbifolded theory $\widetilde{\mathbf{A}}$}

The paired tensor product and (weaker) relative Deligne product can be implemented only from $\mathbf{A}_{\text{ub}}=\overline{\mathbf{A}_{\text{ub}}}$ and its AGQDs. However, it does not directly implement  an interesting \emph{orbifolded} theory (or quasiholes) and \emph{extended}  theory (or electron operators) acting on $\mathcal{H}_{\mathbf{A}}\otimes_{\mathbf{A}_{\text{ub}}}\mathcal{H}_{\mathbf{A'}} $. We stress that the distinction between modules, quasihole operators, and electron operators is fundamental to interpret $\boxtimes_{\mathbf{A}_{\text{ub}}}$. For this purpose, the categorical methods are (rigorous but) complicated and inconvenient at the present stage. Hence, in this subsection, we instead attempt to provide the fusion rule of orbifolded theory $\widetilde{\mathbf{A}}$,

To implement meaningful operators acting on the module $\mathcal{H}_{\mathbf{A}}\otimes_{\mathbf{A}_{\text{ub}}} \mathcal{H}_{\mathbf{A'}}$, often denoted as $\mathbf{A}\boxtimes_{\mathbf{A}_{\text{ub}}} \mathbf{A'}$, one needs to introduce the anomaly-free condition $s_{\alpha_{\text{ub}}}+ s_{\alpha'_{\text{ub}}}=\mathbb{Z}/2$ \cite{Crnkovic:1989ug,Lashkevich:1993fb,Cappelli:1996np,Ino:1998by,Cappelli:2010jv,Gaiotto:2014kfa,Schoutens:2015uia,Zhang:2026gqp,Fukusumi:2026pkr} where $s$ is the conformal spin. This condition is nothing but the boson-fermion spin-statistics relation\cite{Pauli:1940zz}, and one can interpret the object $\alpha_{\text{ub}}\boxtimes \alpha'_{\text{ub}}$ as a (generalized) electron operator\cite{Laughlin:1983fy}. From a technical reason, we clearly distinguish $\alpha_{\text{ub}}\boxtimes_{\mathbf{A}_{\text{ub}}} \alpha'_{\text{ub}}$ from $\alpha_{\text{ub}}\boxtimes \alpha'_{\text{ub}}$. For example, when choosing $\mathbf{A}_{\text{ub}}=\mathbb{Z}_{2}$, the former reduces to the identity, but the latter is still $\mathbb{Z}_{2}$.

We implement the set of operators acting naturally on the paired modules and identify it as an orbifolded theory denoted as $\widetilde{\mathbf{A}}=\left(\mathbf{A}\boxtimes_{\mathbf{A}_{\text{ub}}}\mathbf{A'}\right)/_{\text{ob}}\mathbf{A}_{\text{ub},\boxtimes}$. Then, we introduce the new label $ \widetilde{a}(\in \widetilde{\mathbf{A}})$ as,
\begin{equation}
\mathcal{H}_{(\widetilde{a})}=\sum_{a,a'} M^{\widetilde{a}}_{a,a'}\mathcal{H}_{(a)} \otimes \mathcal{H}_{(a)}
\end{equation}
where $M^{\widetilde{a}}_{a,a'}$ takes $0$ or $1$ and the sum is constrained by the condition $q_{\alpha_{\text{ub}},(a)}=\overline{q_{\alpha'_{\text{ub}},(a')}}$ and the integer spin condition $s_{a}+s_{a'}=s_{\widetilde{a}}\  (\text{mod}.1)$ or the half-integer condition $s_{a}+s_{a'}=s_{\widetilde{a}}\  (\text{mod}.1/2)$ reflecting the type of models. To proceed, we assume that the structure $M^{\widetilde{a}}_{a',a}$  is fixed by some conformal embedding. 

In this setting, one can implement the decomposition of characters and the idempotent $e_{(\widetilde{a})}\in \widetilde{\mathbf{A}}$ as,
\begin{equation}
\Xi_{\widetilde{a}}=\sum_{a,a'} M^{\widetilde{a}}_{a,a'}\chi_{a}  \chi_{a'}, 
\ e_{(\widetilde{a})}=\sum M^{\widetilde{a}}_{a',a}e_{(a)}  e_{(a')}.
\end{equation} 
To proceed, we further assume that AGQDs and idempotents are related to anyons by the Verlinde formula without the $\mathbf{A}_{\text{ub}}$ fixed points. Then the topological operator in $\widetilde{\mathbf{A}}$ is, 
\begin{equation}
\mathcal{Q}_{\widetilde{\alpha}}=\sum_{\widetilde{a},a,a'} q_{\widetilde{\alpha},(\widetilde{a})}M^{\widetilde{a}}_{a,a'}e_{(a)}e_{(a')} 
\label{main_orbifold}
\end{equation}
This form provides a constraint to relate the topological operator in $\widetilde{\mathbf{A}}$ to the original theory $\mathbf{A}\otimes \mathbf{A'}$ through the pairing condition. Focusing the action on the new vacuum $e_{(\widetilde{I})}=\sum_{a}e_{(a)_{\text{ub}}}e_{(a')_{\text{ub}}}$, the necessary condition for $\alpha\boxtimes_{\mathbf{A}_{\text{ub}}}\alpha'\in \widetilde{\mathbf{A}}$ is,
\begin{equation}
\{ q_{\widetilde{\alpha},(\widetilde{I})}=q_{\alpha,(a_{\text{ub}})}q_{\alpha',(a'_{\text{ub}})}\}_{a_{\text{ub}}}
\end{equation}
where $\widetilde{\alpha}$ is a fixed label in $\widetilde{\mathbf{A}}$. 
In the examples in this paper, this condition becomes sufficient, and the following holds,
\begin{equation}
\begin{split}
\widetilde{\mathbf{A}}&=\{\widetilde{\alpha}=\alpha\boxtimes_{\mathbf{A}_{\text{ub}}}\alpha' \in (\mathbf{A}\boxtimes_{\mathbf{A}_{\text{ub}}}\mathbf{A'})_{\text{w}} \\
&:  q_{\widetilde{\alpha},(\widetilde{I})}=q_{\alpha,(a_{\text{ub}})}q_{\alpha',(a'_{\text{ub}})} \text{for all $a_{\text{ub}}=\overline{a'_{\text{ub}}}\in \mathbf{A}_{\text{ub}}$} \}
\end{split}
\label{main_orbifold}
\end{equation}
We conjecture this holds in general and provide a way unifying conformal embedding constructing modular invariants $\sum_{\widetilde{a}}|\Xi_{\widetilde{a}}|^{2}$ (hopefully even with $\mathbf{A}_{\text{ub}}$ fixed points, after proper generalizations). This embedding relates objects in $\mathbf{A}\otimes\mathbf{A'}$ to $\widetilde{\mathbf{A}}$, and this is an example of an anyon condensation preserving the central charge\cite{Cheng:2026qax,Bais:2008ni,Kong:2013aya,Lan:2014uaa,Kong:2017hcw,Zhao:2023wtg}.

\subsection{$\mathbf{A}\boxtimes_{\mathbf{A}_{\text{ub}}} \mathbf{A'}$ as $\mathbf{A}_{\text{ub},\boxtimes}$ extended theory}
Whereas the orbifolded theory is interesting, it does not mean the theory is fundamental in TOs. The puzzle comes from the hidden current $\mathbf{A}_{\text{ub},\boxtimes}$. This object should be identified as a building block of the electron operator in TOs\cite{Milovanovic:1996nj,Cappelli:1996np,Ino:1998by,Cappelli:2010jv,Schoutens:2015uia,Fukusumi:2022xxe,Fukusumi_2022_c,Fukusumi:2024ejk}, whereas $\mathbf{A}_{\text{ub},\boxtimes}$ appears only as hidden currents in the orbifolded model $\widetilde{\mathbf{A}}$. In particular, the new vacuum sector $e_{(\widetilde{I})}=\sum_{a_{\text{ub}}}e_{(a_{\text{ub}})}e_{(a'_{\text{ub}})}$ implies the role of $\mathbf{A}_{\text{ub},\boxtimes}$ under the operator state correspondence. The vertex operators $\Phi_{\alpha_{\text{ub}}\boxtimes \alpha'_{\text{ub}}}$ should be closed under the fusion, and this closed structure should be distinguished from $\Phi_{\alpha_{\text{ub}}\otimes \alpha'_{\text{ub}}}$ in $\mathbf{A}\otimes \mathbf{A'}$. To treat both the electrons $\mathbf{A}_{\text{ub},\boxtimes}$ and the quasiholes $\widetilde{\mathbf{A}}$, one needs to study the \emph{extended} model.

Whereas the hidden current has been studied in CFTs\cite{Crnkovic:1989ug,Lashkevich:1993fb,Gaiotto:2014kfa,Furuta:2025ahl,Zhang:2026gqp,Fukusumi:2026pkr}, the corresponding topological operator is unclear and related aspects have been studied only in \cite{Cordova:2023jip,Albert:2025umy,Cordova:2025eim}. The hidden current should be labelled by hidden operators, outside of $\widetilde{\mathbf{A}}$. We propose the topological operators $\mathbf{A}_{\text{ub},\boxtimes}$ as follows:
\begin{equation}
\mathbf{A}_{\text{ub},\boxtimes}=\{ \mathcal{V}_{\alpha_{\text{ub}}\boxtimes \alpha'_{\text{ub}}}\}_{\alpha_{\text{ub}}},
\mathcal{V}_{\alpha_{\text{ub}}\boxtimes \alpha'_{\text{ub}}}
=\mathcal{Q}_{\alpha_{\text{ub}}\boxtimes_{\mathbf{A}_{\text{ub}}} I} 
\label{main_extension}
\end{equation}
where $\mathcal{Q}_{\alpha_{\text{ub}}\boxtimes_{\mathbf{A}_{\text{ub}}} I} =\sum_{\widetilde{a},a,a'}q_{\alpha_{\text{ub}},(a)} M^{\widetilde{a}}_{a,a'}e_{(a)}e_{(a')} $ holds. $\mathbf{A}_{\text{ub},\boxtimes}$ preserves the fusion rule of $\mathbf{A}_{\text{ub}}$, distinguishing the electron sectors by the topological charge. The preservation of fusion rule $\mathbf{A}_{\text{ub},\boxtimes}\sim \mathbf{A}_{\text{ub}}$ is evident from the appearance of $q_{\alpha_{\text{ub}},(a)}$.  This is the natural extension of the SFC, by inserting $\mathbf{A'}=\overline{\mathbf{A}}$. This reproduces the Verlinde line (topological operator acting on CFTs) $\mathcal{V}_{\alpha\boxtimes\overline{\alpha}}=\sum_{a}q_{\alpha,(a)}e_{(a)}\overline{e_{(a)}}$ in \cite{Petkova:2000ip,Graham:2003nc,Kormos:2009sk}. One can also observe its connection to the modular invariant from $\mathcal{V}_{I\boxtimes \overline{I}}$ by replacing the idempotent $e$ with the character $\chi$. 

Consequently, we propose $
\mathbf{A}\boxtimes_{\mathbf{A}_{\text{ub}}} \mathbf{A'}$ as the (minimum) subalgebra of $(\mathbf{A}\boxtimes_{\mathbf{A}_{\text{ub}}} \mathbf{A'})_{\text{w}}$ described by the extension of $\widetilde{\mathbf{A}}$ by adding objects $\mathbf{A}_{\text{ub},\boxtimes}$. In mathematics, this addition of objects is called \emph{ring adjunction}, and it is denoted as $\widetilde{\mathbf{A}}[\mathbf{A}_{\text{ub},\boxtimes}]$. In short, we propose the following form as an expression of the topological orders,
\begin{equation}
\mathbf{A}\boxtimes_{\mathbf{A}_{\text{ub}}} \mathbf{A'}=\mathbf{A}_{\text{ub},\boxtimes} \times_{\text{ad}} \widetilde{\mathbf{A}}=\widetilde{\mathbf{A}}[\mathbf{A}_{\text{ub},\boxtimes}]
\end{equation}
where we have introduced $\times_{\text{ad}}$ to emphasize the role of extension.
We also expect $\mathbf{A}_{\text{ub},\boxtimes}$-grading structure, $
\mathbf{A}\boxtimes_{\mathbf{A}_{\text{ub}}} \mathbf{A'}=\widetilde{\mathbf{A}}\otimes \mathbf{A}_{\text{ub},\boxtimes}$, for some class of models as a generalization of graded categories in \cite{Turaev:2000ug,etingof2009weaklygrouptheoreticalsolvablefusion,Barkeshli:2014cna,Fukusumi:2024ejk,Fukusumi:2024cnl,Fukusumi:2025ljx}. One can also call $\mathbf{A}\boxtimes_{\mathbf{A}_{\text{ub}}} \mathbf{A'}$ as $\mathbf{A}_{\text{ub},\boxtimes}-$extended symmetry-topological-field theory (SymTFT)\cite{Apruzzi:2021nmk}, in generalized symmetry\cite{Cobanera:2009as,Cobanera:2012dc,Gaiotto:2014kfa}. 

Because we have identified the extended topological operators $\mathbf{A}_{\text{ub}\boxtimes}$, one can straightforwardly construct the corresponding conformal boundary states \cite{Cardy:1986gw,Cardy:1989ir} by using the method in \cite{Graham:2003nc}, This will be interesting in the context of the Li-Haldane conjecture\cite{Li_2008,Qi_2012} and related correspondence\cite{Date:1987zz,Saleur:1988zx,Cardy:2017ufe}. We present the results elsewhere.

\section{Examples of the pairing structures}
\label{section_example}

\subsection{Tricritical Ising and Ising}
We apply our method to the familiar pairing structure appearing in the flow from the tricritical Ising model $\mathbf{M}(4,5)=\{I,\tau\}\otimes \{I,\psi,\sigma\}=\text{Fib}\otimes \text{Ising}$ to the Ising model $\mathbf{M}(3,4)=\{I',\psi',\sigma'\}=\text{Ising'}$, preserving the Ising fusion ring structure\cite{Gaiotto:2012np}. We introduced the Ising fusion ring satisfying $\psi\times \psi=I$, $\psi\times \sigma=\sigma$ and $\sigma\times \sigma=I+\psi$. The pairing is satisfied through the relations $h_{I}+h_{I'}=0$, $h_{\psi}+h_{\psi'}=2$ and $h_{\sigma}+h_{\sigma'}=1/2$. In the following, we denote $y=I,\tau$ to simplify the notation.

$\widetilde{\mathbf{A}}$ is expected to be the Fibonacci fusion ring with,
\begin{equation}
e_{\widetilde{y}}=\sum_{a_{\text{ub}}\in \text{Ising}}e_{(ya_{\text{ub}})}e_{(a'_{\text{ub}})}.
\end{equation}
Hence, the central concern is the underlying modular invariance.
The corresponding partition function $\sum_{\widetilde{y}}|\Xi_{\widetilde{y}}|^{2}$ is calculated as,
\begin{equation}
\sum_{y}|\sum_{a_{\text{ub}}\in \text{Ising}}\chi_{ya_{\text{ub}}}\chi_{a'_{\text{ub}}}|^{2} =\sum_{y}|(\chi_{yI}+\chi_{y\psi})(\chi_{I'}+\chi_{\psi'})|^{2}
\end{equation}
where we have used the triality condition $\chi_{y\sigma}\chi_{\sigma'}=\chi_{yI}\chi_{\psi'}+\chi_{y\psi}\chi_{I'}$ (see Appendix E of \cite{Antunes:2026kqb}), coming from the conformal embedding to $SO(8)_{1}/(G_{2})_{1}$  (see \cite{Gao:2020yep} for related discussions). The right-hand side corresponds to the product of the modular invariant (or the Neveu-Schwarz sectors) of the fermionic minimal models\cite{Petkova:1988cy,Furlan:1989ra,Hsieh:2020uwb}, denoted as $\mathbf{FM}(3,4)\otimes \mathbf{FM}(4,5)$. Hence, we have established the connection between the pairing structure and modular invariance from the following identification at the level of modules,
\begin{equation}
(\mathbf{M}(3,4)\boxtimes_{\text{Ising}}\mathbf{M}(4,5))/_{\text{ob}}\text{Ising}\sim\mathbf{FM}(3,4)\otimes \mathbf{FM}(4,5)
\end{equation}
We expect a similar correspondence between the fermionic models and pairing in the massless flows, as has been noticed in \cite{Crnkovic:1989ug,Lashkevich:1993fb,Gaiotto:2014kfa}.

\section{conclusion}
\label{section_conclusion}
In this paper, we have shown the methods for constructing paired systems 
$(\mathbf{A}\boxtimes_{\mathbf{A}_{\text{ub}}} \mathbf{A'})_{\text{TO}}$ as $\mathbf{A}_{\text{ub},\boxtimes}$ extension of $\widetilde{\mathbf{A}}$, based on the modular data and conformal embedding. This can be regarded as a class of $\mathbf{A}_{\text{ub},\boxtimes}-$extended SymTFT. The hidden current has been identified as generalized electron operators, and this fundametal structure will be useful for further undersatnding of TOs. 

We itemize the open problems:
\begin{itemize}
\item{We focused on the modular invariant, or protected edge modes\cite{Milovanovic:1996nj,Cappelli:1996np,Cappelli:2010jv,2013PhRvX...3b1009L}. However, more generally, the systems can include quasiholes corresponding to (mock) modular covariants. For example, the half-flux quantum in the Moore-Read states corresponds to this structure\cite{Ino:1998by,Fukusumi:2022xxe,Fukusumi_2022_c,Fukusumi:2024ejk}.}
\item{The coherent modular property of $\mathbf{A}\boxtimes_{\mathbf{A}_{\text{ub}}} \mathbf{A'}$ can require the underlying theory $\mathbf{A}$ and/or $\mathbf{A'}$ to be outside of MTCs. In particular, the fixed point property is puzzling even with $\mathbb{Z}_{N}=\mathbf{A}_{\text{ub}}$\cite{Fukusumi:2022xxe,Fukusumi:2023psx,Fukusumi:2024cnl,Fukusumi:2025ljx}. For a general $\mathbf{A}_{\text{ub}}$, the definition of fixed points is unclear. $\mathbf{A}_{\text{ub}}$-grading for $\mathbf{A}$ and $\mathbf{A'}$ will be fundamental, but this property might place the theories outside of MTCs. This seems to be related to the Karoubi envelop\cite{Frohlich:2003hm,Frohlich:2003hg,Gaiotto:2019xmp,Johnson-Freyd:2020usu}. The existing gauging of noninvertible symmetry\cite{Perez-Lona:2023djo,Diatlyk:2023fwf,Perez-Lona:2024sds} will also be helpful for this research direction.}
\end{itemize}

\section{Acknowledgement}
Yoshiki Fukusumi thanks Jean-Emile Bourgine, Satoshi Nakamura, and Yutaka Matsuo for many helpful discussions on the level-rank duality. Yoshiki Fukusumi also thanks Taishi Kawamoto and Shuma Nakashiba for related collaborations. Yoshiki Fukusumi thanks NCTS for support. Yuma Furuta has used generative AI (Anthropic, Fable 5.1) for making figures and free version of Grammarly for improving the texts. The calculation has been performed by ourselves.

\bibliographystyle{ytphys}
\bibliography{Deligne}

\clearpage

\appendix
\section{Examples of the pairing}
\label{example}
\subsection{Group-like pairing in $SU(N)$ model and exceptional Lie group}
In this section, as a simple demonstration of our method, we introduce the results of two series of conformal embeddings to the exceptional Lie groups $(G_{2})_{1} $ and $(F_{4})_{1}$ through the pairing from $\mathbb{Z}_{2}$ and $\mathbb{Z}_{3}$. In general, the $SU(N)_{K}$ Wess-Zumino-Witten model\cite{Wess:1971yu,Witten:1983ar,Witten:1983tw} has $\mathbb{Z}_{N}$ current (or $\mathbb{Z}_{N}$ group-like anyon) $J^{l}$, $L=0,1,...,N-1$, with conformal dimension $h_{J^{l}}=(Kl(N-l))/2N$\cite{Fuchs:1997af}. Hence, $SU(2)_{3}\boxtimes_{\mathbb{Z}_{2}}SU(2)_{1}$ and $SU(3)_{2}\boxtimes_{\mathbb{Z}_{3}}SU(3)_{1}$ are the simplest cases satisfying the paring condition $h_{J^{l}}+h_{J'^{l}}=1$.

In our notation, the corresponding embeddings into the exceptional Lie group in \cite{Schellekens:1986mb,Kac:1988iu,Cordova:2018qvg} are,
\begin{align}
(SU(2)_{3}\boxtimes_{\mathbb{Z}_{2}}SU(2)_{1})/_{{\text{ob}}}\mathbb{Z}_{2}&=(G_{2})_{1} \\
(SU(3)_{2}\boxtimes_{\mathbb{Z}_{2}}SU(3)_{1})/_{{\text{ob}}}\mathbb{Z}_{3}&=(F_{4})_{1}
\end{align}
One can straightforwardly confirm the above identification from the preservation of the Fibonacci fusion ring (see \cite{Lin:2023uvm} for example) and the termination of the $\mathbb{Z}_{2}$ or $\mathbb{Z}_{3}$ structure. For further studies, we note that the $\mathbb{Z}_{N}$ pairing conditions for the $SU(N)_{K}$ have already been classified in \cite{Fukusumi:2026tii}, and related anomaly-free subgroup structures have been studied in \cite{Closset:2025lqt}.

\subsection{$SU(N)_{K}\boxtimes SU(K)_{N}=SU(NK)_{1}$ duality}
The level-rank duality $SU(N)_{K}\otimes SU(K)_{N}\sim SU(NK)_{1}$\cite{Kuniba:1990zh,Nakanishi:1990hj,Hsin:2016blu,Cordova:2017vab,Cordova:2023jip,Bourgine:2024ycr} is the most familiar examples which involves with the pairing structure. We concentrate our attention on the setting with $\text{gcd}(N,K)=1$ to avoid puzzles from fixed points\cite{Han:2017hdv,Numasawa:2017crf}. In this setting, the decompotion $SU(N)_{K}=\mathbb{Z}_{N}\otimes \mathbf{A}_{\text{ub}}$ and 
 $SU(K)_{N}=\mathbb{Z}_{K}\otimes \mathbf{A'}_{\text{ub}} $ hold. We conjecture the equivalence $\mathbf{A}_{\text{ub}}=\overline{\mathbf{A}_{\text{ub}}}$ and the resultant expression,
\begin{equation}
SU(N)_{K}\boxtimes_{\mathbf{A}_{\text{ub}}} SU(K)_{N}=\mathbf{A}_{\text{ub},\boxtimes}\times_{\text{ad}}SU(NK)_{1}
\end{equation}  
Hence, $\mathbf{A}_{\text{ub},\boxtimes}$ extension of $SU(NK)_{1}$ will realize TO with generalized electrons with nonabelian fusion rule $\mathbf{A}_{\text{ub},\boxtimes}$. 
We also conjecture the following decomposition, or $M^{\widetilde{I}}$, 
\begin{equation}
e_{\widetilde{I}}=\sum_{a_{\text{ub}} \in \mathbf{A}_{\text{ub}}} e_{(a_{\text{ub}})}e_{(a'_{\text{ub}})}
\end{equation}
This expression clarifies $\mathbf{A}_{\text{ub},\boxtimes}$ as uncharged particles, belonging to the new vacuum $\widetilde{I}$.
By applying the $\mathbb{Z}_{N}\otimes \mathbb{Z}_{K}=\mathbb{Z}_{NK}$ structure, one will directly obtain the desired transformation law for $SU(NK)_{1}$. By further stacking  $U(1)_{NK}$ with $\mathbb{Z}_{NK}$ to the system, one can construct $GL(NK)_{1}$ fermion model with $\mathbf{A}_{\text{ub},\boxtimes}\otimes \mathbb{Z}_{NK}$ electron\cite{Nakanishi:1990hj}. This explains the existence of non-group-like symmetry in the fermionic systems\cite{Wei:2026fsn}.

For example, the $SU(3)_{2}$ and $SU(2)_{3}$ model share a Fibonacci anyons, $\tau$ or $\tau'$, and they satisfy the relation $h_{\tau}+h_{\tau'}=1$. Because of this pairing, only the $\mathbb{Z}_{2}\otimes \mathbb{Z}_{3}=\mathbb{Z}_{6}$ structures survive after the orbifolding, and this perfectly matches $SU(6)_{1}$. 
We only note the vacuum sectors of the orbifolded theory,
\begin{equation} 
e_{(\widetilde{I})}=e_{(I)}e_{(I')}+e_{(\tau)}e_{(\tau')}
\end{equation}
One can distinguish the sector $e_{(I)}e_{(I')}$ and $e_{(\tau)}e_{(\tau')}$ by acting $\mathcal{V}_{\tau\boxtimes \tau'}$, through the AGQDs $q_{\tau, (\ )}=(1\pm\sqrt{5})/2$.
By applying $\mathbb{Z}_{6}=\mathbb{Z}_{2}\otimes \mathbb{Z}_{3}$ structures, one can obtain $6$ idempotents. Consequently, the following expression holds,
\begin{equation}
SU(3)_{2}\boxtimes_{\mathbf{A}_{\text{ub}}} SU(2)_{3}=\{I\boxtimes I',\tau \boxtimes \tau'\}\otimes \mathbb{Z}_{6} 
\end{equation}

\section{Massless renormalization group flow and symmetry enforcement in fine tuning}
In this section, we note a concise way relating RG domain wall and pairing structure. The massless RG is an RG flow from an ultraviolet (UV) CFT $\mathbf{A}\boxtimes \overline{\mathbf{A}}$ to the infrared (IR) CFT $\overline{\mathbf{A'}}\boxtimes \mathbf{A'}$ under some relevant perturbation $\Phi_{\text{rel}}$. For simplicity, we drop $\boxtimes$ in this section. One can express the phenomenon by the corresponding Hamiltonian $H$ as,
\begin{equation}
H_{\mathbf{A}}+\lambda\Phi_{\text{rel}} \Rightarrow H_{\mathbf{A'}}
\end{equation}
where $\lambda$ is the coupling constant, $\Rightarrow$ is mapping between theories. In \cite{Brunner:2007ur,Gaiotto:2012np}, the problem has been mapped to the conformal interface problem. Related references usually focus on the Lagrangian formalism. We attempt to provide a concise argument by using the quantum Hamiltonian and demonstrate a massless RG as a spin- and $\mathbf{A}_{\text{ub}}$-preserving interface.

We focus on the three points,
\begin{itemize}
\item{Perturbative expressions of Hamiltonian QFTs\cite{Cardy:1986ie,Yurov:1989yu,Yurov:1991my} should be consistent.}
\item{$\Phi_{\text{rel}}$ should be compatible with the symmetry $\mathbf{A}_{\text{ub}}$.}
\item{The spin of $\Phi_{\text{rel}}$ is $0$.}
\end{itemize}
The last point has been regarded as granted, but it is fundamental to see the naturalness of the framework.

First, let us focus on a UV (nonchiral) primary state $|a\rangle$. Applying the perturbation, this state changes to $|a\rangle_{\lambda}$. Hence, if the system flows to $\mathbf{A'}$, the following identification should hold.
\begin{equation}
|a\rangle_{\lambda}\sim |x'\rangle
\end{equation}
where $|x'\rangle$ is some IR state.
Then, we apply the spin $0$ property of $\Phi_{\text{rel}}$. Because the spin of  $\Phi_{\text{rel}}$ is 0, the spin $s$ of the quantum states should be preserved.
\begin{equation}
s_{a}=s_{x'}
\end{equation}
This is nothing but the preservation of $L_{0}-\overline{L_{0}}$ where $L_{n}$ is the Virasoro generators. Hence, one can \emph{naturally (or naively)} introduce the generalization of this spin preservation,
\begin{equation}
(L_{n}-\overline{L_{n}})|a\rangle_{\lambda}\sim  (L'_{n}-\overline{L'_{n}})|x'\rangle
\end{equation}
When interpreting the mapping as an interface, this is nothing but a conformal interface. In other words, the conformal interface should be regarded as a particular type of spin-preserving interface, which appears rigorously. We expect that this restricition from the spin-preserving interface to the conformal interface has some connection to the integrability of the systems.

Next, we study the unbroken $\mathbf{A}_{\text{ub}}(=\overline{\mathbf{A'}_{\text{ub}}})$ symmetry. Because the unbroken symmetries are conserved charges, the following should hold,
\begin{equation}
q_{\alpha_{\text{ub}},(a)}|a\rangle_{\lambda}=\mathcal{Q}_{\alpha_{\text{ub}}}|a\rangle_{\lambda}\sim\overline{\mathcal{Q}_{\alpha'_{\text{ub}}}}|x'\rangle
\end{equation}
Because of the above equality and correspondence, $\overline{\mathcal{Q}_{\alpha'_{\text{ub}}}}|x'\rangle$ should be expressed as,
\begin{equation}
\overline{\mathcal{Q}_{\alpha'_{\text{ub}}}}|x'\rangle=\overline{q_{\alpha'_{\text{ub}},(a')}}|x'\rangle,
\end{equation}
where $a'$ is an IR primary field constrained by the pairing condition or superselection rule $q_{\alpha_{\text{ub}},(a)}=\overline{q_{\alpha'_{\text{ub}},(a')}}$.
Hence, by applying the folding trick and focusing on the chiral parts, $|a\rangle_{\lambda} \otimes \overline{|x'\rangle}$ produces the pairing $\otimes_{\mathbf{A}_{\text{ub}}}$.

We comment on the discrepancy between the analytical formulations and the RG domain wall. The point is that unbroken $\mathbf{A}_{\text{ub}}$ symmetry does \emph{not} ensures the absense of the IR relevant fields from $\Phi_{\text{rel}}$\cite{Fukusumi:2026pkr}. Hence, to realize $\mathbf{A'}$ in a stable way, one needs to enforce additional symmetries $\mathbf{O}$ in general. However, one cannot enforce $\mathbf{O}$ freely, because $\mathbf{O}\times_{\text{ad}} \mathbf{A}_{\text{ub}}$ pairing structure, $\mathcal{H}_{\mathbf{A}}\otimes_{\mathbf{O}\times_{\text{ad}} \mathbf{A}_{\text{ub}}} \mathcal{H}_{\mathbf{A'}}$, can break the orginal $\mathbf{A}_{\text{ub}}$ pairing structure $\mathcal{H}_{\mathbf{A}}\otimes_{\mathbf{A}_{\text{ub}}} \mathcal{H}_{\mathbf{A'}}$. This incompatibility may imply the illdefinedness of the original $\mathbf{A}_{\text{ub}}$ preserving interface or that of $\mathbf{O}\times_{\text{ad}} \mathbf{A}_{\text{ub}}$ preserving interface as a realization of the massless RG $\mathbf{A}\Rightarrow \overline{\mathbf{A'}}$. When the $\mathbf{O}\times_{\text{ad}} \mathbf{A}_{\text{ub}}$-preserving interface becomes ill-defined in this context, it might signal the existence of intermediate IR theories $\mathbf{A}_{\text{int}}$ between $\mathbf{A}$ and $\mathbf{A'}$. Generally, the larger symmetry $\mathbf{O}\times_{\text{ad}} \mathbf{A}_{\text{ub}}$ constrains the form of the relevant perturbation $\Phi_{\text{rel}}$ more strongly, and this constraint permits the flow to the unexplored intermediate theory $\mathbf{A}_{\text{int}}$. In other words, the flow $\mathbf{A}\Rightarrow \overline{\mathbf{A'}}$ is prohibitted through the symmetry $\mathbf{O}\times_{\text{ad}}\mathbf{A}_{\text{ub}}$ from the structure of modules.  The nontrivial fixed points in \cite{Dotsenko:2006eq,Dotsenko:2007sh,Estienne:2010zza} may correspond to $\mathbf{A}_{\text{int}}$. We note the similarity between this consistency of the pairing and the symmetry span in \cite{Ando:2026ffy}.

Finally, we note the possible obstacle for the RG domain wall. The spin-preserving interface is \emph{weaker} than a conformal interface, and one cannot exclude the possibility of a massless RG outside of the RG domain wall at this stage. Moreover, it has been established that the pairing condition does \emph{not} match the 't Hooft or LSM anomaly matching condition\cite{Kikuchi:2022ipr,Herviou:2023unm,Fukusumi:2026pkr}. Hence, whereas RG domain walls should provide massless RGs canonically, the inverse is uncertain.

\section{Gapped domain wall from $\widetilde{\mathbf{A}}$}
In the main text, we focused on the following reduction,
\begin{equation}
\widetilde{\mathbf{A}}\subset \mathbf{A}\boxtimes_{\mathbf{A}_{\text{ub}}}\mathbf{A'}\subset  (\mathbf{A}\boxtimes_{\mathbf{A}_{\text{ub}}}\mathbf{A'})_{\text{w}}\subset \mathbf{A}\otimes \mathbf{A'}
\end{equation}
One can see the relationship evidently from the expressions by idempotents.

One can also obtain the mapping $\mathbf{A}\rightarrow \overline{\mathbf{A'}}$ from $\widetilde{\mathbf{A}}$. The procedure is simple by interpreting the idempotent as a dual vector,
\begin{equation}
e_{(\widetilde{a})}=\sum_{a,a'}M^{\widetilde{a}}_{a,a'}e_{(a)}e_{(a')}\rightarrow e_{(\widetilde{a}),\mathbf{A}^{\dagger}}= \sum_{a,a'}M^{\widetilde{a}}_{a,a'}e^{\dagger}_{(a)}\overline{e_{(a')}}
\end{equation}
where $e^{\dagger}_{(a)}$ is the dual vector, satisfying the contraction rule $\mu(e_{a}^{\dagger},e_{(b)})=\delta_{a,b}$. Hence, the application of $\rho=\mu ( e_{(\widetilde{a}),\mathbf{A}^{\dagger}}, \ )$ to $\mathbf{A}$ maps object in $\mathbf{A}$ to $\overline{\mathbf{A'}}$. One can freely choose the $\mathbb{C}$ linear combination $\rho=\sum_{\{\widetilde{a}\}=\mathbb{S}_{\times}} C_{\widetilde{a}} \mu (e_{(\widetilde{a}), \mathbf{A}^{\dagger}}, \ )$ and construct the corresponding domain walls where $C$ is the coefficients and $\mathbb{S}_{\rho,\times}$ is the subset of label of primary fields in $\widetilde{\mathbf{A}}$. By carefully choosing $\mathbb{S}_{\rho,\times}$ and $C_{\widetilde{a}}=1$, one can obtain an identity-preserving mapping, so-called a monoidal functor or ring homomorphism. The preservation of the identity requires one to choose the appearance of all of the idempotents $\overline{e_{(a')}}$ and $e_{(a)}^{\dagger}$ once with the coefficient $1$ in the linear sum of $\rho$. This argument provides a simple construction of gapped domain walls\cite{Kitaev:2011dxc,Lan:2014uaa,Kong:2017hcw,Zhao:2023wtg}, respecting the pairing. Because the mapping relation of the idempotent has been completely included in $\mathbb{S}_{\rho,\times}$, the result reproduces the analysis in \cite{Fukusumi:2025clr,Fukusumi:2025xrj,Fukusumi:2025fvb}. This construction of a monoidal functor from the relative Deligne product seems to correspond to the Eilenberg-Watts-type calculus in \cite{Fuchs2016EilenbergWattsCF,Fuchs2020ModuleEC}.

We focused on the mapping structure from $\widetilde{\mathbf{A}}$, but one can freely attach the $\mathbf{A}_{\text{ub}}=\overline{\mathbf{A'}_{\text{ub}}}=\mathbf{A}_{\text{ub},\boxtimes}$ structure to the mapping when assuming the corresponding grading structures of $\mathbf{A}$, $\mathbf{A'}$ and $\mathbf{A}\boxtimes_{\mathbf{A}_{\text{ub}}}\mathbf{A'}$. This is $\mathbf{A}_{\text{ub}}$ enrichment, and this extension naturally introduces the charged domain walls\cite{Fukusumi:2024ejk,Fukusumi:2025ljx,Fukusumi:2026pkr}.

\end{document}